# Temperature dependence of $Er^{3+}$ ionoluminescence and photoluminescence in $Gd_2O_3$:Bi nanopowder

Zuzanna Boruc,[1,*], Grzegorz Gawlik, [2] Bartosz Fetliński,[1]
Marcin Kaczkan,[1] Michał Malinowski[1]

[1]*Institute of Microelectronics and Optoelectronics, Warsaw University of Technology, ul. Koszykowa 75, 00-662 Warsaw, Poland*
[2]*Institute of Electronic Materials Technology, ul. Wolczynska 133, 01-919 Warsaw, Poland*

**Abstract**

Ionoluminescence (IL) and photoluminescence (PL) of trivalent erbium ions ($Er^{3+}$) in $Gd_2O_3$ nanopowder host activated with $Bi^{3+}$ ions has been studied in order to establish the link between changes in luminescent spectra and temperature of the sample material. IL measurements have been performed with $H_2^+$ 100 keV ion beam bombarding the target material for a few seconds, while PL spectra have been collected for temperatures ranging from 20 to 700°C. The PL data was used as a reference in determining the temperature corresponding to IL spectra. The collected data enabled the definition of empirical formula based on the Boltzmann distribution, which allows the temperature to be determined with a maximum sensitivity of $9.7 \times 10^{-3}$ °C$^{-1}$. The analysis of the $Er^{3+}$ energy level structure in terms of tendency of the system to stay in thermal equilibrium, explained different behaviors of the lines intensities. This work led to the conclusion that temperature changes during ion excitation can be easily defined with separately collected PL spectra. The final result, which is empirical formula describing dependence of fluorescence intensity ratio on temperature, raises the idea of an application of method in temperature control, during processes like ion implantation and some nuclear applications.

## I. INTRODUCTION

The ionoluminescence (IL) is a type of luminescence, in which excitation energy is delivered to the material by an incident ion beam. The IL method has been well known for many years [1-3]. However, it was only in the 90's that information first became available on the IL of rare earth elements [4]. Agullo et al. [5] proposed a diagram presenting events occurring after ion beam bombardment, while Calderon et al. [6] extended a description of possible relaxation paths of energy deposited in a material. Researchers are still finding new applications of IL [7]. E.g., recently, the variation of IL spectra was used for detection of proton dose in quartz [8].

The essence of the IL phenomenon is energy deposition within a material after an ion went through it. The energy delivered to the system causes ionization of the material atoms and results in the creation of an electron cloud. The energy is then passed on to the crystal lattice and its impurities. Basically, energy can be directed to create phonons, which causes a rise of the sample temperature, or to form luminescent centers. Thus, after ion bombardment, two processes can be observed, namely the heating of a material and the emission of light – the ionoluminescence.

[*]Author to whom correspondence should be addressed. Electronic mail: z.boruc@stud.elka.pw.edu.pl

The luminescence is strictly dependent on impurities and defects present in the material. Luminescent spectra change with temperature, which affects the distribution of electrons between energetic levels potentially involved in the act of emission. This behavior is used in the Fluorescence Intensity Ratio (FIR) method. FIR technique is based on the variation of the intensity ratio between two thermally coupled emitting energy levels.

As demonstrated in this paper, data obtained from both luminescence phenomena, IL and photoluminescence (PL), can be used to determine the temperature of a material during ion beam excitation. As IL is part of many industrial processes, such measurements may be highly useful during ion implantation, where limiting the temperature is essential due to defects induced in silicon crystal, or in temperature control in nuclear installations. Temperature in such installations is currently measured with thermocouples, which have a big disadvantage in that they are introduced physically, so even the slightest defect in the assembly system can lead to false results or even a radioactive leak.

In this paper, we propose a general idea of a new approach to measurements of temperature in systems where IL occurs.

**II. DESCRIPTION OF EXPERIMENT**

Gadolinium oxide ($Gd_2O_3$) doped with 1 at.% Er and 1 at.% Bi (hereafter named $Gd_2O_3$:Er,Bi) nanopowder was synthesized using the sol-gel method [9] in the Institute of Electronic Materials Technology in Warsaw. The resulting grain size was about 80 nm. $Gd_2O_3$, a host similar to the widely known $Y_2O_3$, is characterized by good thermal stability, chemical durability and a low phonon cutoff frequency of about 600 cm$^{-1}$. $Gd^{3+}$ ions can easily be substituted by rare earth (RE) ions, e.g. erbium ion, because the atomic radius is similar.

The experimental part of this research was twofold. In the first part, IL spectra of $Gd_2O_3$:Er,Bi were measured, in the second, PL spectra for the same sample were collected for temperatures ranging from room temperature up to 800 °C. The variation of temperature for PL measurements was chosen to make these results comparable to IL results, where sample heating was induced by an ion beam. Such a combination of PL and IL measurements, allowed calculating temperature that occurred during ion irradiation.

During IL measurements, a 1 mm thick layer of $Gd_2O_3$:Er,Bi nanopowder was placed in the ion implanter chamber, on a small aluminum tray covering the area of about 5 mm x 5 mm. The target material was exposed to a molecular hydrogen ($H_2^+$) swift ion beam, with energy of 100 keV and a current density of approximately 3.3 µA/cm$^2$. The ion beam footprint covered whole sample area, however, only the first layer of the target material was irradiated, as the experiment was conducted in a vacuum chamber of an ion implantation system. The IL signal was measured using a fiber optic waveguide and minispectrometer C10083CAH, manufactured by Hamamatsu and equipped with a back-thinned type CCD camera with a spectral response from 320 to 1000 nm. The measurement system enabled recording 100 luminescent spectra in 10 seconds and thus the system resolution was 0.1 s. One-minute breaks were introduced to let the sample and a holder cool to avoid influence of temperature gain during previous run.

During the PL experiment, the emission spectra were measured using a Digikrom 480 grating monochromator manufactured by the CVI Laser Corporation, followed by a WCT 50 photomultiplier tube manufactured by Thorn EMI and a SR400 gated photon counter manufactured by Stanford Research Systems. The sample was excited with a 355 nm wavelength obtained by the third harmonic of a pulsed Continuum Surelite Nd:YAG laser (10 ns pulse-width, repetition rate 10 Hz). The irradiated area of sample material was equal to a laser beam footprint with diameter of approximately 3 mm. Sample heating was provided by a compact heater Superthal Mini MS 31 manufactured by Kanthal.

### III. RESULTS AND DISCUSSION

**A. Ionoluminescence spectra**

This experiment consisted of 23 runs of ion beam irradiation. Each run was 10 seconds long and was conducted with approximately 1 minute breaks of ion beam operation between runs. Three of the selected IL spectra recorded during the first ion beam run are presented on Fig. 1. The series of sharp lines in a green band is attributed to the erbium ions and a violet broad band is caused by bismuth ions. The energy of the ion beam is used for excitation of electrons (inelastic interactions) and creation of point defects (elastic interactions). The distribution of energy between those two interactions, can be described by the $S_e/S_n$ ratio, where $S_e$ is the electronic stopping power and $S_n$ is the nuclear stopping power. Both parameters have been calculated with SRIM software [10]: $S_e$ = 194 keV/(mg/cm$^2$) and $S_n$ = 0,87 keV/(mg/cm$^2$). As the $S_e/S_n$ ratio is equal to 223, it means that most of the deposited energy is consumed on electronic effects rather than nuclear ones. This explains the presence of a strong luminescence excited by energy from the ion beam. As the $S_e$ value is much higher than $S_n$, which is typical for swift ions, the nuclear effects can be neglected in further investigations.

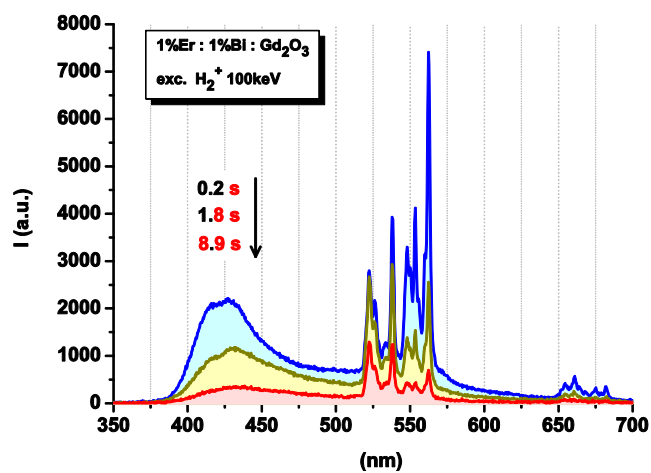

**FIG. 1.** IL spectra of Gd$_2$O$_3$:Er,Bi recorded during first ion beam run. The three spectra were recorded after 0.2, 1.8 and 8.9 seconds of ion beam irradiation.

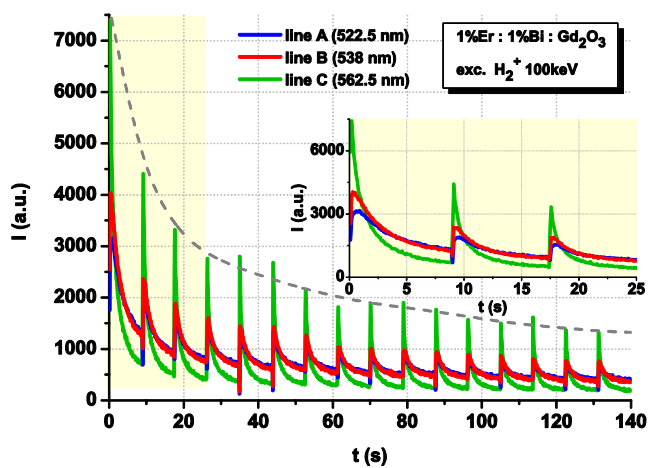

**FIG. 2.** The IL intensity maxima of Er$^{3+}$ ions emission lines during time of hydrogen ion beam irradiation. Peaks on the curves are preceded by 1 minute breaks of ion beam irradiation. Breaks are not indicated on the timescale. The inset shows the IL spectra of the first three runs.

After each cooling step, all emission lines tended to return to their initial intensities. However, maximal intensity achieved at the beginning of each run decreased with each consecutive run, see dotted line on Fig. 2. This could be explained by a model of IL degradation caused by ion beam irradiation as proposed by Gawlik et al. in [11], which is based on the assumption that observed IL signal decrease is caused by ion beam generation of competitive nonradiative recombination centers. Such point defects in our case can be created at the end of the stopping process, when energy of ions drops to values of a few keV and the stopping occurs by elastic interaction. The final effect of this is a stable point defect. Previously to IL experiments with Gd$_2$O$_3$:Er:Bi powder, there were several IL measurements conducted in the laboratory with various crystalline bulk materials. With bulk crystalline materials, we were able to detect defects by using the RBS channeling method. Initial conclusions were that the diminishing of the signal may be caused by the accumulation of point defects. However, it should be mentioned that the nature of these defects is not clearly understood and is still under investigation.

A final observation was the fast, reversible decrease in IL intensity during each run. This is attributed to an increase in material temperature due to ion beam irradiation. However, the behavior of line A and C strongly differs, especially during the first second of each run, see inset in Fig. 2. The intensity of line A initially slightly grows and then decreases. Luminescence of line C constantly diminishes. The most pronounced effect is a dramatic change of proportion between line A and C during a single ion beam run, see Fig. 3. The proportion C/A of intensities of lines C and A appears to depend on temperature but also to be insensitive to ion fluence. The difference in maximum values appears to be within measurement error. In all runs, the C/A ratio changes from a value of about 3.5 for a room temperature sample to 0.5 for the highest temperature sample. This effect seems to be independent of ion beam degradation of the IL signal. Thus, the ratio of Er$^{3+}$ IL lines can be used for determining the temperature of the irradiated material.

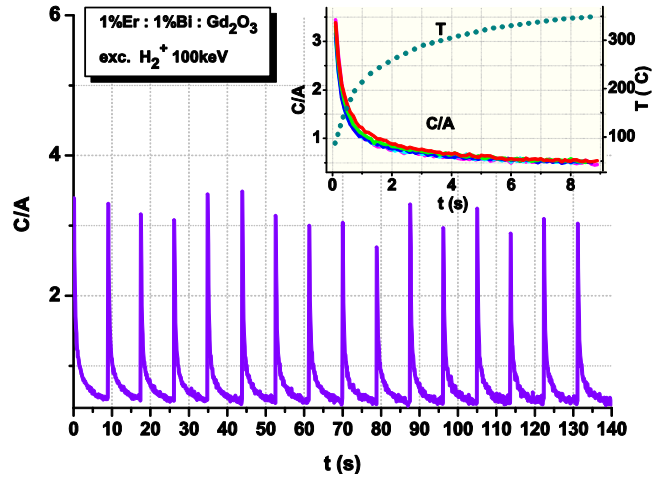

**FIG. 3.** Proportion of $Er^{3+}$ IL emission lines marked C and A in Fig. 1. Inset shows a ratio of intensities C/A for $1^{st}$, $2^{nd}$, $3^{rd}$, $8^{th}$, and $15^{th}$ run of ion beam.

In summary, two types of changes of IL spectra were observed during ion beam irradiation. Firstly, a slow persistent diminishing of IL intensity due to material degradation by the ion beam (see dotted line on Fig. 2) and secondly, a faster IL intensity decrease, resulting from material heating caused by energy deposited in the material by ion beam (see solid lines on Fig. 2).

In order to explain the reason for the phenomena described above, the PL spectra were measured for a broad range of temperatures, see part B of Section III.

### B. Photoluminescence spectra

The $Er^{3+}$ energy levels structure in $Gd_2O_3$ nanocrystals has already been thoroughly investigated by Chen et al. [12] and is presented in a simplified version in Fig. 4. The photoluminescence characteristics of energy transfer between $Er^{3+}$ and $Bi^{3+}$ ions in the same host are described by Xu et al. in [13].

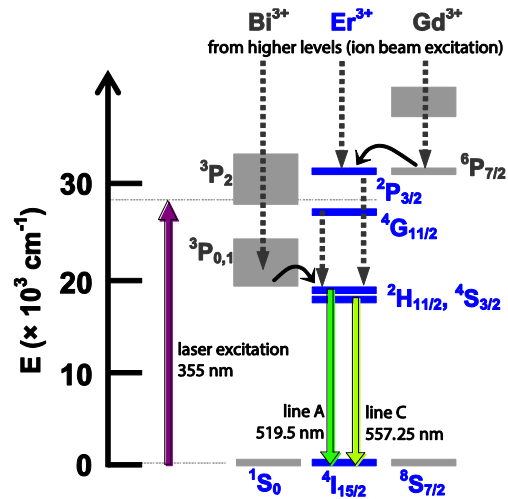

**FIG. 4.** Simplified energy levels scheme of $Er^{3+}$, $Gd^{3+}$ and $Bi^{3+}$ ions in $Gd_2O_3$ host. The dotted arrow lines correspond to nonradiative transitions.

It was shown recently by Ju et al. [14] that $Bi^{3+}$ ions usually enhance the $Ln^{3+}$ emission due to strong broadband absorption of $Bi^{3+}$ in the region between 300 and 400 nm. The absorption of exciting photons is followed by the efficient energy transfer from $Bi^{3+}$ to $Ln^{3+}$ ions. Yang et al. in [15] studied an influence of $Bi^{3+}$ ions on $Er^{3+}$ ions in a $Y_2O_3$ host and showed that the presence of $Bi^{3+}$ ions enhances the intensity of the $Er^{3+}$ emission for a 330 nm excitation of approximately 42 times. For the $Gd_2O_3$ host, the maximum intensity in the excitation spectrum was centered around 340 nm [13]. Thus, the chosen excitation line at 355 nm seems to be a reasonable choice for our sample.

Emission spectra obtained for temperatures varying from 23 to 800 °C are shown in Fig. 5. Three emission lines are considered, A = 519.5 nm, B = 534 nm and C = 557.25 nm which correspond to $(^2H_{11/2}, ^4S_{3/2}) \rightarrow ^4I_{15/2}$ transitions. The intensities of the lines change with temperature in accordance with the Boltzmann distribution, as the relative intensities grow with temperature for higher lying levels and decrease for lower lying levels. This rule is not only valid for distribution of population between multiplets but also between Stark components of each multiplet. This was confirmed by the fact that not only does the A/C ratio (A, C – emissions from different multiplets) grow with temperature but the A/B ratio also does (A, B – emissions from the same multiplet). This is caused by a phenomenon of thermal excitation and results from the tendency of the system to stay in thermal equilibrium.

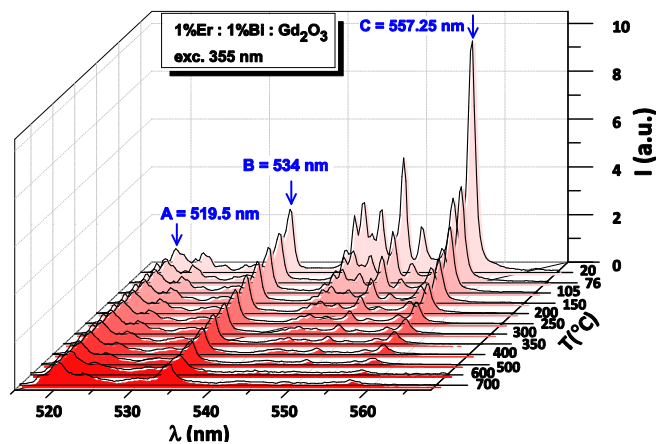

**FIG. 5.** Temperature dependence of the spectra collected in the 20 °C – 700 °C range. Emission at 519.5 nm (line A) is normalized to unity.

The energy gap between the lowest Stark level of $^2H_{11/2}$ multiplet and the highest Stark level of $^4S_{3/2}$ is about 240 cm$^{-1}$ at room temperature. Such energy difference is small enough to consider both levels as being thermally coupled. The measure of probability of transferring an electron to a higher energy state is an expression based on the Boltzmann distribution exp(-ΔE/kT). This means that when kT is equal to ΔE, the probability of transferring an electron to a higher state is equal to 0.37. The kT value changes from 200 cm$^{-1}$ for room temperature up to 680 cm$^{-1}$ for 700°C. The kT value is equal to 240 cm$^{-1}$ for 70 °C.

The process of transferring part of the population of the level to the higher state is depicted on Fig. 6, which shows how intensities of considered lines change with growing temperature. It can be seen that while the intensity of line C constantly falls, the intensities for line A and B grow in the same range. This is the consequence of the $Er^{3+}$ energy levels structure, where it is not possible for $^4S_{3/2}$ level to be thermally populated by any lower lying level. The population of $^4S_{3/2}$ decreases with rising temperature. Just above the $^4S_{3/2}$, the $^2H_{11/2}$ multiplet is located, which population initially grows with temperature. This is caused by the thermal excitation of electrons from state $^4S_{3/2}$ to state $^2H_{11/2}$.

The spectrum in Fig. 5 for 700 °C demonstrates that almost the whole population of $^4S_{3/2}$ multiplet has been transferred to the $^2H_{11/2}$ multiplet. Such a significant change in the distribution of the $^2H_{11/2}$, $^4S_{3/2}$ population can be used for temperature measurements, widely known as phosphor thermometry [16].

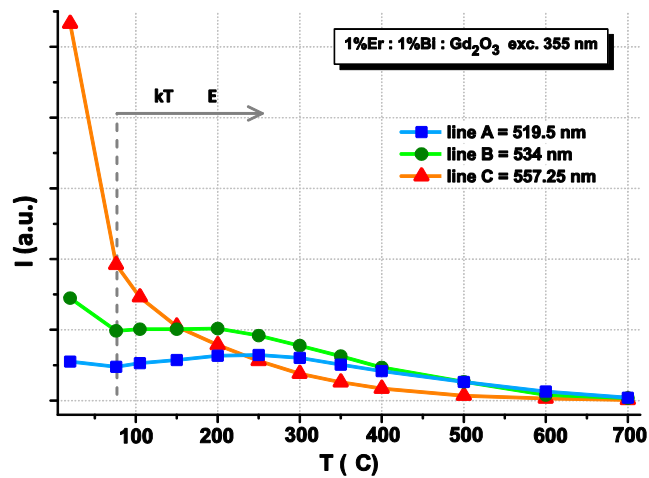

**FIG. 6.** Temperature dependence of the emission intensity of three considered lines. A, B and C.

To obtain a mathematical formula which would associate relative intensity with temperature, the appropriate fitting of data has to be conducted. As previously mentioned, the thermal coupling between considered levels should follow the Boltzmann distribution:

$$N_1/N_2 = C_{RD} \times \exp(-\Delta E/kT) \qquad (1)$$

where $N_1$, $N_2$ are the populations of considered energy levels, $C_{RD}$ is the coefficient, which depends on the radiative transition rates and the degeneration of these levels and T is the temperature in Kelvins. When T is the variable in the above expression, the resulting sensitivity is dependent on the ΔE value. The pair of lines A and C was the used in further calculations.

The best representation of data exhibiting exponential dependence is the Arrhenius plot, which allows the setting of data linearly, as it uses a logarithmic scale for the intensity ratio and an inverse scale for temperature. The ratio A/C of intensity values at 519.5 nm (line A, 19 249 cm$^{-1}$) and 557.25 nm (line C, 17 945 cm$^{-1}$) is plotted against the inverse temperature in Fig. 7. The obtained slope value of 1 155 cm$^{-1}$ is close to the theoretical one of 1 304 cm$^{-1}$. The difference may result from other energy transfer processes present in the investigated system, which is not described here. The probability of the occurrence of such processes is often dependent on thermal excitation, as the energy delivered with temperature can easily fill the mismatch in possible cross-relaxation channels.

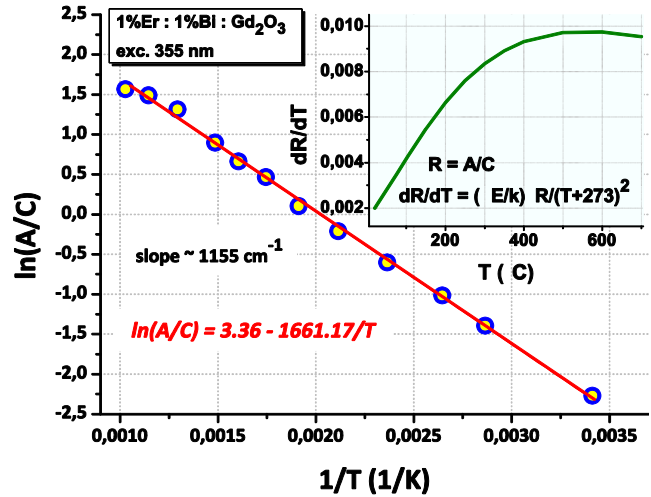

**FIG. 7.** Arrhenius plot of the A and C lines emissions intensity ratio temperature dependence. The inset shows resulting sensitivity.

The obtained empirical formula is:

$$\ln(A/C) = 3.36 - 1661.17/T \qquad (2)$$

To define the resulting sensor sensitivity, the differential of the above expression should be calculated: $dR/dT = (dE/k) \times R/T^2$, where R is the A/C ratio. The resulting sensitivity is presented in the inset in Fig. 7. The range of the highest sensor sensitivity is between 500 and 600 °C, where dR/dT reaches the value of $9.7 \times 10^{-3}$, which is a relatively high value compared to typical results obtained in phosphor thermometry using other rare earth ions. In [17] results for up-converted $Er^{3+}$ emissions in $Gd_2O_3:Yb^{3+}$ were presented with a maximum sensitivity of $3.9 \times 10^{-3}$ for 300 K. In [18] the maximum sensitivity of $5.1 \times 10^{-3}$ for 495 K was obtained for $Er^{3+}$ in $Al_2O_3:Yb^{3+}$, which is similar to our results. Although we cannot compare directly our results with both of those cited here, it seems that using $Er^{3+}$ ions in crystals provide high sensitivities in temperature measurements and that the presence of $Bi^{3+}$ ions strongly enhances $Er^{3+}$ emission.

The derived formula (2) allows us to calculate the temperature of the material, knowing only the intensity ratio of two lines: $T = 1661.17/(3.36-\ln(A/C))$. This formula was used to calculate the temperature in the inset in Fig. 3. It may seem unusual, that the initial temperature of each run is not equal to room temperature, but it should be remembered that the integration time for each IL spectrum was equal to 0.1 s, which corresponds to a maximum resolution of collected spectra. Thus, it is quite understandable that the initial C/A ratio may not be the ratio at room temperature, but a slightly higher temperature.

## C. Convergence of IL and PL spectra

The nature of the observed decrease of IL signal with temperature is clearly twofold. There are obviously two different processes responsible for the diminishing of luminescence.

The first process is not reversible, the degradation is persistent, and it slowly progresses with each consecutive run. It is caused by ion beam degradation of the sample material and is highlighted in Fig. 2 with a dotted line. The model explaining this phenomenon is based upon the idea of generation of competitive nonradiative centers by implanted ions with some efficiency and preservation of existing luminescent centers [11]. However, it seems difficult to apply this model for approximation of degradation caused by ion beam for thermally coupled levels.

The second process is reversible. The luminescence quickly decays during ion bombardment, but restores initial intensity at the beginning of each record. As there were 1 minute intervals between each record to let the sample cool, this effect can be associated with temperature. Moreover, during our IL experiments the $Er^{3+}$ spectra were changing slowly, which proves that the rise of temperature was originating from thermal vibrations rather than energy of overlap of atom in a crystal structure. It was shown in part B of Section III how levels $^2H_{11/2}$ and $^4S_{3/2}$ are thermally coupled, which sufficiently explains this reversible degradation as well as the different character of intensity decreases for each multiplet.

The independence of both degradation processes is shown in Fig. 3 where it can be seen that the initial and final values of relative intensities (ratio A/C), as well as the character of decay, stay the same for each run of the experiment.

The main result of comparison of both IL and PL spectra is a determination of sample temperature during ion bombardment. This can be performed with Eq. (2). The observed change in the intensity ratio for IL corresponds to the temperature rise from approximately 80 to 350 °C during each run.

Finally, attention should be given to the initial part of each run. In the inset in Fig. 2 (IL) it can be seen that the intensities of line A and B are growing during the first second of run. This is confirmed by Fig. 6 (PL), where it can be seen that those intensities start to grow just above 75-80°C (when kT value exceeds ΔE), which is equal to the temperature of each run after about 0.1 s.

## IV. CONCLUSIONS

We present here a method for measuring the temperature of a material during ion bombardment. The idea is based on the FIR method, where intensities of two thermally coupled emission lines are compared. The first step was the calibration with use of a PL spectra measured for different temperatures of a sample. Intensity ratio of the 520 and 557 nm emission lines changes accordingly to the Boltzmann distribution, giving a linear dependence on logarithmic plot. This enabled establishment of the empirical formula which describes the relation between intensity ratio and temperature. The resulting maximum sensitivity is high and equals $9.7 \times 10^{-3}$ $K^{-1}$. The second step was the application of the obtained formula to determine the temperature during ion beam excitation of the sample. The presented experiment is based upon the assumption that the Boltzmann distribution is valid regardless of the method of excitation. This is true, provided that no energy transfer process exists in the examined system, such as up-conversion.

Another significant goal of this investigation, was to show how incorporation of two luminescent methods can provide deeper and better understanding of the observed processes than each of these methods on its own. For example, without PL temperature dependent data, it would not be possible to determine the nature of changes like different decaying of the considered emission lines or the presence of rising time for some emission lines during ion excitation.

Finally, the method presented here may potentially be used for in situ control of the temperature, e.g. during ion implantation, when temperature rise must be limited so that the target material structure is not destroyed, or in nuclear applications, as it enables contactless, wireless measurements, and thus can provide a safer way to control temperature of the parts subjected to energetic particle irradiation.